\documentclass[12pt]{article}
\pdfoutput=1
\usepackage{setspace}
\usepackage{amsmath}
\usepackage{amssymb}
\usepackage{colortbl}
\usepackage[table,x11names]{xcolor}
\usepackage{graphicx}
\usepackage{hyperref}
\usepackage{cite}
\usepackage{slashed}
\usepackage{pifont}
\usepackage{longtable}
\usepackage{multirow}
\usepackage{ltablex}

\definecolor{nicered}{rgb}{0.6,0,0}
\definecolor{nicegreen}{rgb}{0.1,0.5,0.1}
\definecolor{niceblue}{rgb}{0,0.4,0.8}
\hypersetup{colorlinks,citecolor= nicegreen,linkcolor=
  nicered,urlcolor=niceblue}

\allowdisplaybreaks

\begin{document}
\begin{titlepage}
% -------------------------------------------------
% Uncomment to add preprint numbers when necessary
% -------------------------------------------------
% \begin{flushright}

% \end{flushright}
  \newcommand{\AddrLiege}{{\sl \small IFPA, D\'ep. AGO, Universit\'e de
      Li\`ege, B\^at B5, Sart Tilman B-4000 Li\`ege 1,
      Belgium}}
  \newcommand{\AddrUFSM}{{\sl \small  Universidad T\'ecnica Federico Santa Mar\'{i}a,
      Departamento de F\'{i}sica\\Casilla 110-V, Avda. Espa\~na 1680, Valparaiso, Chile}}
  \vspace*{0.5cm}
\begin{center}
  \textbf{\Large Closing in on minimal dark matter and\\[3mm]
    radiative neutrino masses}
  \\[9mm]
  D. Aristizabal Sierra$^{a,b,}$\footnote{e-mail address: {\tt
      daristizabal@ulg.ac.be}},
  C. Simoes$^{a,}$\footnote{email address: {\tt csimoes@ulg.ac.be}},
  D. Wegman$^{a,}$\footnote{email address: {\tt dwegman@ulg.ac.be}}
  \vspace{0.8cm}\\
  $^a$\AddrLiege\\[7mm]
  $^b$\AddrUFSM\\[9mm]
\end{center}
\vspace*{0.2cm}
\begin{abstract}
  \onehalfspacing
  We study one-loop radiative neutrino mass models in which one of the
  beyond-the-standard model fields is either a hypercharge-zero
  fermion quintet (minimal dark matter) or a hypercharge-zero scalar
  septet. By systematically classifying all possible one-loop such
  models we identify various processes that render the neutral
  component of these representations (dark matter) cosmologically
  unstable. Thus, our findings show that these scenarios are in
  general not reconcilable with dark matter stability unless tiny
  couplings or additional ad hoc symmetries are assumed, in contrast
  to minimal dark matter models where stability is entirely due to the
  standard model gauge symmetry. For some variants based on
  higher-order loops we find that $\alpha_2$ reaches a Landau pole at
  rather low scales, a couple orders of magnitude from the characteristic scale of the
  model itself. Thus, we argue that some of these variations although
  consistent with dark matter stability and phenomenological
  constraints are hard to reconcile with perturbativity criteria.
\end{abstract}
\end{titlepage}
\setcounter{footnote}{0}
% -----------------
% Introduction
% -----------------
\section{Introduction}
\label{sec:intro}
Neutrino physics, dark matter (DM) and the cosmic baryon asymmetry
provide strong evidence for beyond-the-Standard Model (BSM)
physics. Neutrino masses and their mixing pattern can be
well-described at the effective level by a certain
lepton-number-violating operator \cite{Weinberg:1979sa}, with the
dim=5 being the obvious choice. Conventional wisdom is that this
operator originates from a type-I seesaw involving a GUT scale. The
reason for this is probably related with the fact that such picture
nicely fits within $SO(10)$-inspired scenarios, while at the same time
explains the absence of charged lepton-flavor-violating signals in
high-intensity experiments \cite{Agashe:2014kda} \footnote{Seesaw-like
  scenarios with sizeable lepton-flavor-violating effects can be
  constructed, but they require deviations from the ``standard''
  type-I seesaw picture, see e.g. \cite{Sierra:2012yy}.}. In contrast
to neutrino physics, there is no such ``standard'' scenario for DM,
and several phenomenologically-motivated frameworks exist
\cite{Feng:2010gw}. However, of particular popularity are WIMP
scenarios where the DM relic density results from thermal freeze-out.

Other well-motivated scenarios for neutrino mass generation exist and
provide a potential link between the origin of neutrino masses and
that of DM \cite{Ma:2006km} (see e.g. \cite{Sierra:2008wj} for a
phenomenological analysis). This connection somehow resembles what one
finds in the type-I seesaw, where neutrino masses and the cosmic
baryon asymmetry find a common explanation through standard
leptogenesis (see \cite{Davidson:2008bu,Fong:2013wr} for reviews)
\footnote{Variants of the standard leptogenesis picture can be
  embedded within type-I seesaw or in other neutrino mass models as
  well, see
  e.g. \cite{AristizabalSierra:2007ur,AristizabalSierra:2009bh}.}. Particularly
relevant for that endeavor are models where the dim=5
lepton-number-violating operator arises through radiative corrections
(see \cite{Bonnet:2012kz,Sierra:2014rxa} for a full list of all such
possibilities at the one- and two-loop level). In these contexts
several approaches have been adopted to assure DM stability. The
conventional one relies on the introduction of ad hoc symmetries under
which the SM and the dark sector transform differently, thus
guaranteeing stability. These symmetries can have multiple origins,
and can be regarded as remnants of the spontaneous symmetry breaking
of a more fundamental symmetry related with e.g. GUTs
\cite{Kadastik:2009cu,Frigerio:2009wf} or flavor theories
\cite{Sierra:2014kua}.

Another approach that has been explored relates radiative neutrino
masses with higher-order electroweak (EW) representations
\cite{Cai:2011qr,Kumericki:2012bh,Kumericki:2012bf,Law:2013saa,Chen:2011bc,Cai:2016jrl,Ahriche:2015wha,Culjak:2015qja,Ahriche:2016rgf}.
In this case the idea is different and relies on the mechanism
underlying minimal DM models
\cite{Cirelli:2005uq,Cirelli:2007xd,Cirelli:2009uv}, namely for
higher-order EW representations renormalizable interactions with SM
operators cannot be written, and so they are absolutely stable at the
renormalizable level (at the renormalizable level the minimal DM
Lagrangian exhibits an accidental $\mathbb{Z}_2$ symmetry). Decays are
induced by effective operators, thus in these scenarios DM is never
absolutely stable but can be cosmologically stable provided the
effective operator is sufficiently suppressed. In
\cite{Cirelli:2005uq} it was shown that the stability condition, the
requirement of EW gauge coupling perturbativity up to
$M_\text{Planck}$ and direct DM searches constraints, singles out two
possible representations: hypercharge-zero fermion quintet and
hypercharge-zero scalar septet, of which the fermion quintet defines
the minimal DM model. Recently, in refs. \cite{DiLuzio:2015oha} and
subsequently in \cite{DelNobile:2015bqo} it was shown that for the
scalar septet one can write a loop-induced lower dimensional operator
that renders the septet a non-viable DM representation, even if the
effective scale is assumed to be $M_\text{Planck}$. Thus, within the
minimal DM context the only viable representation is the fermion
quintet, although recently analyses have shown that depending on the
DM profile this representation is not consistent with indirect DM
searches results \cite{Cirelli:2015bda,Garcia-Cely:2015dda} (see
discussion in sec. \ref{sec:DM-decay}).

In this paper we show that when the fermion quintet acts as a mediator
in one-loop neutrino mass generation, there are always DM decay
operators that---under reasonable parameter choices---lead to fast DM
decays, regardless of the one-loop neutrino mass model. Some of these
operators arise at the tree level but others are loop-induced. The
latter being specially important in models where in addition to the
fermion quintet, representations beyond quartets are present. We will
show that as soon as the minimal DM scheme is extended to include
other representations (that allow the construction of one-loop
neutrino mass models) there are always $\mathbb{Z}_2$-breaking
couplings, thus fast decay modes are always expected to be present. We
consider the case of the scalar septet as well, despite being not not
consistent with stability even in the minimal DM framework. The reason
is that these results allow the identification of DM fast decay modes
induced by neutrino physics itself, rather than due to a different
type of physics (related with unknown quantum gravity effects).

The rest of this paper is organized as follows. In
sec. \ref{sec:DM-decay}, we briefly review minimal DM models,
introduce our notation and present formulas that will be used in our
estimations. In sec. \ref{sec:dm-decay-processes}, we present the
different $\mathbb{Z}_2$-breaking decay operators and estimate
lifetimes for various processes, showing that even for neutrino
physics parameters taken to their extreme values the decays are always
fast. In sec. \ref{sec:UV-completions}, we systematically derive all
one-loop realizations of the Weinberg operator, under the condition of
the UV completion containing either the fermion quintet or scalar
septet. With these results at hand, we then identify the different DM
decay operators associated with each of the neutrino mass model
categories. In sec.~\ref{sec:possible-caveats}, we discuss the main
pitfalls of models for Majorana neutrino masses with higher-order EW
representations and briefly comment on models with higher-order
loops. Finally, in sec. \ref{sec:concl} we summarize and present our
conclusions.
% -----------------
% Section 1
% -----------------
\section{Minimal dark matter and possible decay processes}
\label{sec:DM-decay}
Minimal DM models rely on the observation that for higher-order
$SU(2)$ representations, $\boldsymbol{R}$, renormalizable gauge
invariant operators of the form
\begin{equation}
  \label{eq:effective-operator-DM}
  \mathcal{O}_{N=4}\sim \boldsymbol{R}\,\mathcal{O}_\text{SM}\, ,
\end{equation}
where $\mathcal{O}_\text{SM}$ is an operator involving just SM fields,
cannot be written
\cite{Cirelli:2005uq,Cirelli:2007xd,Cirelli:2009uv}. Thus, when the SM
is endowed with such state, its stability is automatically
guaranteed. The stabilization mechanism at work, resembles the SM
``mechanism'' which assures proton stability, namely the gauge
symmetry does not allow for renormalizable operators that might induce
decay. Of course, at the non-renormalizable level several effective
operators of the form (\ref{eq:effective-operator-DM}) can be written,
but their renormalizable realizations require degrees of freedom which
the minimality criteria do not allow for \footnote{Something of this
  sort is found in the SM as well. Baryon- and lepton-number-violating
  effective interactions can be written
  \cite{Weinberg:1979sa,Wilczek:1979hc}, but their renormalizable
  forms require beyond-the-SM physical states.}.

The lowest-order representation for which the above argument proves to
be true and yields a viable DM candidate is the fermionic
$\boldsymbol{R}=\boldsymbol{5}$. Let us discuss this in more
detail. For $\boldsymbol{R}=\boldsymbol{3}$, renormalizable operators
can always be written, regardless of whether they are fermions ($F$)
or scalars ($S$) and provided $Y=0,1$. For the quartet and beyond it
is more subtle. Using the notation $\boldsymbol{R}_X^Y$ (with $X$
referring to fermion ($F$) or scalar ($S$) and $Y$ to hypercharge,
normalized according to: $Q=T_3+Y$), and bearing in mind that
\begin{equation}
  \label{eq:su2-decompositions}
  \underbrace{\boldsymbol{2}\otimes\dots \otimes 
    \boldsymbol{2}}_{(R-1)\;\;\text{times}}\supset \boldsymbol{R}\, ,
\end{equation}
one finds that for the quartet non-renormalizable operators are
possible depending on its spin, namely
\begin{equation*}
  \label{label:fourplet-decomposition}
  \mathcal{O}_{N=5}\sim \boldsymbol{4}_F^Y\,\mathcal{O}_\text{SM}
  \,
  \left\{
    \begin{aligned}
      \mathcal{O}_\text{SM}&=\ell\,H\,H^\dagger\quad (Y=1/2)\\
      \mathcal{O}_\text{SM}&=\ell\,H^\dagger\,H^\dagger\quad (Y=3/2)
    \end{aligned}\, , \right.  \,
  \mathcal{O}_{N=4}\sim \boldsymbol{4}_S^Y\,\mathcal{O}_\text{SM} \,
  \left\{
    \begin{aligned}
      \mathcal{O}_\text{SM}&=H^\dagger\,H\,H^\dagger\quad (Y=1/2)\\
      \mathcal{O}_\text{SM}&=H^\dagger\,H^\dagger\,H^\dagger\quad (Y=3/2)
    \end{aligned}
  \right.\, .
\end{equation*}
However, though involving non-renormalizable operators,
$\boldsymbol{R}_F=\boldsymbol{4}_F$ does not provide a viable DM
candidate (the $\boldsymbol{R}_S=\boldsymbol{4}_S$ involves
renormalizable couplings and so it is of course non-viable
either). First of all, even if the cutoff scale is taken to be
$\Lambda=M_\text{Planck}$ (where arguably quantum gravity effects will
generate such operator) its suppression is not sufficiently strong to
guarantee cosmological stability. Furthermore, in the context of
minimal DM only $Y=0$ representations are consistent with direct DM
searches: For $Y\neq 0$ states, $Z$-mediated processes induce a
tree-level spin-independent DM-nucleon cross section well above the
values allowed by current DM direct searches
\cite{Cirelli:2005uq,Cirelli:2007xd,Cirelli:2009uv}. We stress that
the validity of this statement is subject to the minimality condition.
Departures from this assumption, e.g. by introducing another state
that mixes with the DM producing a small mass splitting, enable
$Y\neq 0$ representations (for more detailes see
\cite{Cirelli:2009uv}). For $\boldsymbol{R}_F=\boldsymbol{5}_F$ one
finds
$\boldsymbol{2}\otimes \boldsymbol{2}\otimes \boldsymbol{2}\otimes
\boldsymbol{2}\supset \boldsymbol{5}$,
which then implies that $SU(2)$ invariance requires four SM
fields. Depending on the spin of the representation, the effective
operator in (\ref{eq:effective-operator-DM}) is $N=5$ (scalar) or
$N=6$ (fermion), with its full structure determined by the hypercharge
of the corresponding state.

\begin{figure}
  \centering
  \includegraphics[scale=1.2]{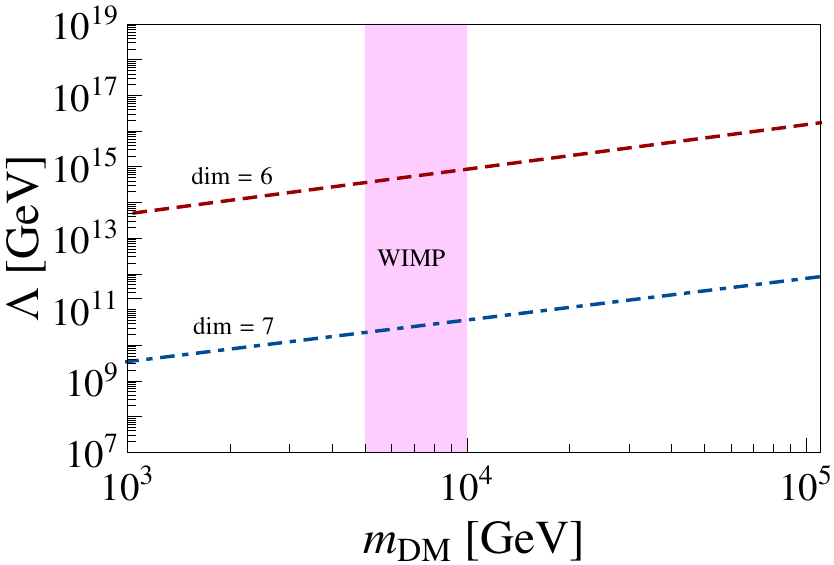}
  \caption{Contour lines of constant
    $\tau_\text{DM}=10^{26}\,$~seconds as a function of the cutoff
    scale $\Lambda$ and $m_\text{DM}$. From top to bottom the contour
    lines refer to dimension 6 and 7 effective operators that induce
    DM decay. The vertical stripe encloses the DM mass region where
    the DM relic density results from thermal freeze-out.}
  \label{fig:lifetimes-minimal-DM-only}
\end{figure}

An upper limit on the size of the representations can be determined by
the condition of $\alpha_2$ being perturbative up to the
Planck scale. Ref. \cite{DiLuzio:2015oha} has shown that depending on
the representation a precise determination of the Landau pole should
rely on two-loop RGE. Based on that analysis representations
$\boldsymbol{R}_F>\boldsymbol{5}_F$ and
$\boldsymbol{R}_S>\boldsymbol{8}_S$ are ruled out, as has been
previously pointed out in
\cite{Cirelli:2005uq,Cirelli:2007xd,Cirelli:2009uv} using a one-loop
RGE calculation. One is left then with
$\boldsymbol{R}<\boldsymbol{5}_{F,S}$ and $\boldsymbol{7}_S$ for which
the decay lifetimes can be estimated from the effective
operator
\begin{equation}
  \label{eq:eff-operator}
  \mathcal{O}_N=\frac{c_N}{\Lambda^{N-4}}
  \,\boldsymbol{R}\mathcal{O}_\text{SM}\, .
\end{equation}
where the Wilson coefficient $c_N$, assumed to be order one for
simplicity, can involve flavor indices too.  In the limit
$m_\text{DM}\gg v$ (with
$v=\langle H\rangle=(\sqrt{2}G_F)^{-1/2}\simeq 246\,$~GeV), the total
decay width can be parameterized as follows \cite{DiLuzio:2015oha}
\begin{equation}
  \label{eq:total-decay-width}
  \Gamma_\text{DM}^{(N)}=\frac{1}{4(4\pi)^{2N-5}}\,\frac{m_\text{DM}^{2N-2n_c-7}}
  {(N-2)!(N-3)!}\frac{(v/\sqrt{2})^{2n_c}}{\Lambda^{2(N-4)}}\, ,
\end{equation}
where $n_c$ refers to the number of Higgs condensate insertions.  Of
the three potentially viable representations only those for which
$\tau_\text{DM}\gtrsim 10^{26}\,$~sec (as required by indirect
detection experiments looking for $\gamma$-ray, $\nu$, $e^+$ or $p^-$
signals stemming from DM decay
\cite{Ando:2015qda,Rott:2014kfa,Ibarra:2013zia,Giesen:2015ufa}) are
consistent. From (\ref{eq:total-decay-width}) their lifetimes can be
calculated. Consistency with $\tau_\text{DM}\gtrsim 10^{26}\,$~sec for
$\boldsymbol{5}^0_S$ requires scales way above $M_\text{Planck}$, in
contrast $\boldsymbol{5}^0_F$ and $\boldsymbol{7}^0_S$ have
sufficiently long lifetimes even for scales well below the Planck
scale, due to their decays being driven by dim=6 and dim=7 operators
as can be seen in fig. \ref{fig:lifetimes-minimal-DM-only}.

In the absence of any other degree of freedom, these results hold if
no lower dimension operator for these representation can be
written. For $\boldsymbol{5}^0_F$ this proves to be true, for
$\boldsymbol{7}^0_S$, instead, not. The point is that the previous
analysis rely on tree level effective decay modes, but for
$\boldsymbol{7}^0_S$ one-loop radiative dim=5 operators can be written
through the operator
$\boldsymbol{7}^0_S
\left(\boldsymbol{7}^0_S\boldsymbol{7}^0_S\right)H^\dagger HH^\dagger
H$
\cite{DiLuzio:2015oha} (see discussion below). Thus, this argument
rules out the $\boldsymbol{7}^0_S$ as well.  Minimal DM therefore
reduces to a single possibility, $\boldsymbol{5}^0_F$, for which
stringent constraints from indirect DM searches have been found. Of
particular relevance are those coming from $\gamma$-ray line searches
from the galactic center, for which it has been proven that if the
Milky Way possesses a Navarro-Frenk-White or Einasto DM profiles this
representation is ruled out too
\cite{Cirelli:2015bda,Garcia-Cely:2015dda} \footnote{We thank Thomas
  Hambye for pointing this out to us.}. It can be however consistently
considered in the context of cored profiles such as Burket or
Isothermal.
\subsection{Minimal DM decay in the presence of extra degrees of
  freedom}
\label{sec:dm-decay-processes}
Minimal DM instability is induced by UV completions associated with
effective operators of the type (\ref{eq:eff-operator}). In the
minimal model the new degrees of freedom are assumed to be related
with quantum gravity effects, something that guarantees stability for
$\boldsymbol{5}^0_F$. For UV completions associated with ``lighter''
physics, DM can become cosmologically unstable. For
$\Lambda\lesssim 10^{15}\,$~GeV, $\boldsymbol{5}^0_F$ has a lifetime
below $10^{26}\,$~sec.

The neutrino mass one gets from one-loop realizations of the Weinberg
operator can be estimated as
\begin{equation}
  \label{eq:neu-mm-one-loop}
  m_{\nu}\sim \frac{v^2}{16\pi^2}\frac{Y^4}{\Lambda}\, ,
\end{equation}
where $Y$ denotes a generic Yukawa coupling. Thus, assuming
$Y\subset [10^{-2},1]$ and fixing the neutrino mass to
$m_\text{Atm}=0.05\,$~eV
\cite{Forero:2014bxa,Gonzalez-Garcia:2014bfa,Capozzi:2013csa}, one can
estimate $\Lambda \lesssim [10^5,10^{13}]$~GeV. Note that in this
estimation the one-loop function has been neglected, but in an
specific model with an explicit neutrino mass matrix the presence of
the loop function will lead to smaller cutoff scales. Thus, with the
value derived from (\ref{eq:neu-mm-one-loop}) we are overestimating
the DM lifetime.

This result then shows that---in principle---any one-loop UV
completion of the Weinberg operator that can yield in turn an UV
completion for the effective operator responsible for
$\boldsymbol{5}^0_F$ decays will lead to:
\begin{equation}
  \label{eq:life-time}
  \tau_\text{DM}\lesssim 2\times 10^{18}\,
  \left(\frac{10^4\,\text{GeV}}{m_\text{DM}}\right)^5\,
  \left(\frac{\Lambda}{10^{13}\,\text{GeV}}\right)^4\,\text{sec.}\, ,
\end{equation}
and therefore to unviable ``minimal DM''\footnote{Strictly speaking
  once minimal DM is embedded in radiative neutrino mass models (or
  any other scenario) is not anymore minimal DM since its defining
  conditions do not hold anymore.}. This conclusion can be understood
as follows. In the absence of further degrees of freedom the minimal
DM full Lagrangian reads:
\begin{equation}
  \label{eq:full-Lag}
  \mathcal{L}=\mathcal{L}_\text{SM} 
  + \text{Tr}\left[\overline{\boldsymbol{5}_F}\,i\slashed{D}\,\boldsymbol{5}_F
  \right]
  + \frac{1}{2}m_5
  \left(
    \overline{\boldsymbol{5}_F^c}\,\boldsymbol{5}_F
    +
    \text{H.c.}
  \right)\, .
\end{equation}
This Lagrangian is invariant under the $\mathbb{Z}_2$ transformations
$\boldsymbol{5}_F\to - \boldsymbol{5}_F$ and
$X_\text{SM}\to X_\text{SM}$, that entirely results as a consequence
of the gauge symmetry.  If the extra representations that allow for
the one-loop neutrino mass matrix allow as well for UV completions of
the following Lagrangian
\begin{equation}
  \label{eq:5-eff}
  \mathcal{L}_\text{eff} = \frac{c_6}{\Lambda^2}
  \,\boldsymbol{5}_F\,\ell\,H\,H^\dagger\,H\, ,
\end{equation}
they will necessarily involve interactions that explicitly break the
accidental $\mathbb{Z}_2$ symmetry (as this operator does) and
therefore will lead to DM decay processes with typical lifetimes given
by (\ref{eq:life-time}). Instability of the $\boldsymbol{7}_S^0$ can
be understood in the same way. The full Lagrangian is given by
\begin{equation}
  \label{eq:lag-seven*}
  \mathcal{L}= 
  \left|D_\mu\,\boldsymbol{7}_S\right|^2 + V_\text{SM} + 
  m_7^2|\boldsymbol{7}_S|^2 + \lambda_1|\boldsymbol{7}_S|^2|H|^2
  + \cdots\, ,
\end{equation}
where the ellipses refer to additional terms, that as the ones we have
explicitly written are $\mathbb{Z}_2$-invariant. In this case the
relevant effective $\mathbb{Z}_2$-breaking term is
\begin{equation}
  \label{eq:eff-7}
  \mathcal{L}_\text{eff}=
  \frac{c_5}{\Lambda}\,\boldsymbol{7}_S\,\boldsymbol{7}_S
  \,\boldsymbol{7}_S\,H^\dagger\,H\, .
\end{equation}
This operator alone cannot yield DM decays, but when combined with the EW
invariant $H^\dagger\,H$ leads to a one-loop dim=5 decay operator
\cite{DiLuzio:2015oha}. As in the fermionic DM case, instability of
the $\boldsymbol{7}_S$ can be understood as a consequence of
renormalizable $\mathbb{Z}_2$-violating interactions that in the
effective limit reduce to (\ref{eq:eff-7}).

As already pointed out, decay modes depend upon the extra
representations and can be sorted according to tree level and one-loop
induced decay modes. In what follows whenever calculating lifetimes we
will set $m_\text{DM}=10^4\,$~GeV and the mass of the extra
representations, assumed to be universal ($m$), to $10^9\,$~GeV, which
is typically the largest value for the extra degrees of freedom one
will get in a specific neutrino mass model. With the relevant masses
fixed in this way, the $\mathbb{Z}_2$-conserving Yukawa couplings
(those related with neutrino mass generation: $Y_\nu$) are fixed to
one \footnote{This parameter choice is a simplification that enables
  determining the largest lifetimes achievable in each case. In
  practice, however, these scales are not entirely independent and
  instead are constrained by phenomenological conditions such as the
  DM relic abundance.}.  We start our analysis with the tree level
modes:
\begin{itemize}
\item \textbf{Tree level decay modes for $\boldsymbol{5}^0_F$}\\
  These DM decay processes are mediated by scalar or fermion $Y=1/2$
  quartets and sextets, as required by gauge invariance. In both cases
  one can distinguish processes with no additional representations
  (single messenger scenarios) and processes with further
  representations, e.g. scalar triplets or any other enabled by gauge
  invariance (multiple messenger scenarios). In single messenger
  processes, DM decays proceed via off-shell states that in turn
  directly decay to SM particles, through either renormalizable or
  non-renormalizable couplings (examples are shown in
  fig. \ref{fig:tree-level}, diagrams $(a), (c), (d)$).  In multiple
  messenger scenarios, instead, DM decay via cascade processes
  mediated by several off-shell states, see fig. \ref{fig:tree-level}
  diagram $(b)$ for an example. Note that one can as well write
  multiple messenger decay modes for the sextet by expanding the
  effective couplings $c_6^S$ and $c_7^F$ in diagrams $(c)$ and $(d)$
  in fig.~\ref{fig:tree-level}. In what follows we discuss in more
  detail the examples shown in fig. \ref{fig:tree-level}, aiming at
  showing that in models involving quartets tree level DM decay
  processes always lead to fast DM decay, while for those with sextets
  these processes do not lead to rapid decay modes, therefore
  motivating the investigation of DM decay loop-induced processes.

  Of particular relevance for one-loop neutrino mass models is
  $\boldsymbol{4}^{1/2}_S$ for which one can write
  $\boldsymbol{5}^0_F\to \ell\,\boldsymbol{4}^{1/2}_S\to
  \ell\,H\,H^\dagger\,H$
  (see fig. \ref{fig:tree-level}).  Depending on the different $SU(2)$
  contractions this operator chain contains several decay processes,
  of which $l^\pm\,W_L^\mp\,Z_L\,Z_L$ is the dominant decay mode. The
  lifetime can be estimated by inserting $c_6=Y\cdot\lambda$ in
  eq. (\ref{eq:total-decay-width}) and trading $\Lambda$ to $m$. The
  result reads:
  \begin{equation}
    \label{eq:lifetime-four-messenger}
    \tau_\text{DM}\lesssim 2\times 10^6\,
    \left(\frac{10^{-2}}{\lambda}\right)^2
    \left(\frac{10^4\,\text{GeV}}{m_\text{DM}}\right)^5\,
    \left(\frac{m}{10^9\,\text{GeV}}\right)^4
    \,\text{sec.}\, ,
  \end{equation}
  which rules out the possibility of stable
  $\chi_\text{DM}\subset \boldsymbol{5}^0_S$ in one-loop neutrino mass
  models that contain a $\boldsymbol{4}^{1/2}_S$ (see discussion in
  sec. \ref{sec:possible-caveats}). As an example of multiple
  messenger processes we consider $\boldsymbol{4}^{1/2}_F$ and
  $\boldsymbol{3}^0_S$, tree level decay processes associated with
  $\boldsymbol{5}^0_F\to H^\dagger\,\boldsymbol{4}^{1/2}_F\to
  H^\dagger\,\bar\ell\,H\,H^\dagger$
  can be written. In contrast to the single messenger case, in these
  scenarios as obviously expected more $\mathbb{Z}_2$-violating
  couplings are involved (see fig. \ref{fig:tree-level}). The lifetime
  can be estimated according to
  \begin{equation}
    \label{eq:lifetime-multiple-mess}
    \tau_\text{DM}\lesssim 2\times 10^{18}\,
    \left(\frac{10^{-2}}{h}\right)^2
    \left(\frac{10^3\,\text{GeV}}{\mu}\right)^2
    \left(\frac{10^4\,\text{GeV}}{m_\text{DM}}\right)^5\,
    \left(\frac{m}{10^9\,\text{GeV}}\right)^6
    \,\text{sec.}\, .
  \end{equation}
  Here $h$ refers to a generic Yukawa coupling (see
  fig. \ref{fig:tree-level}, diagram $(b)$).
  \begin{figure}[t!]
    \centering
    \includegraphics[scale=0.55]{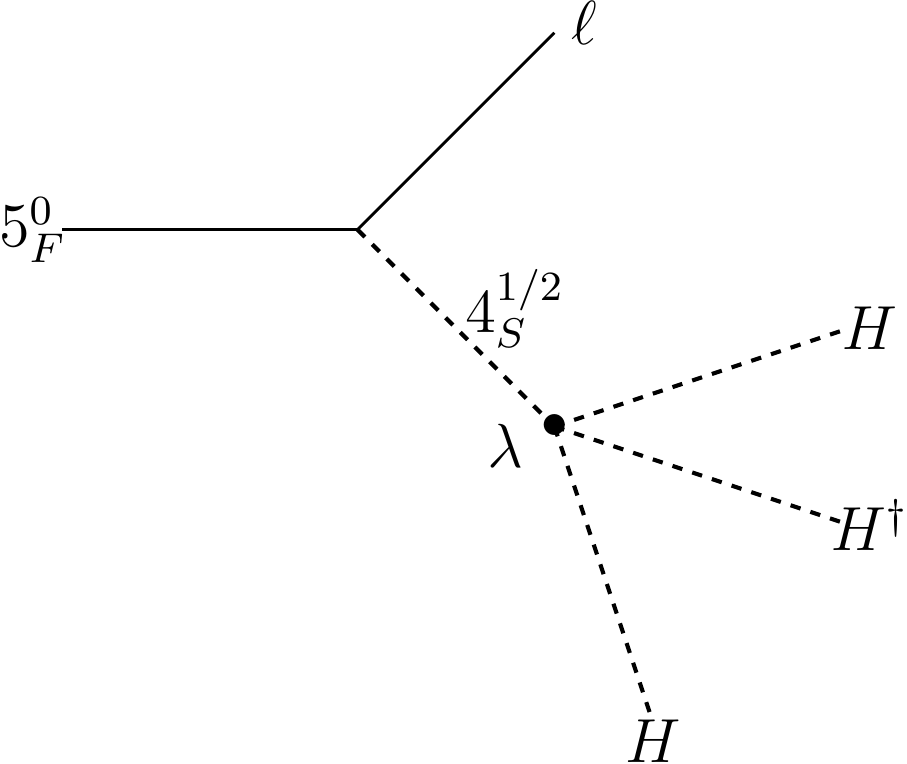}
    \hspace{2.5cm}
    \includegraphics[scale=0.55]{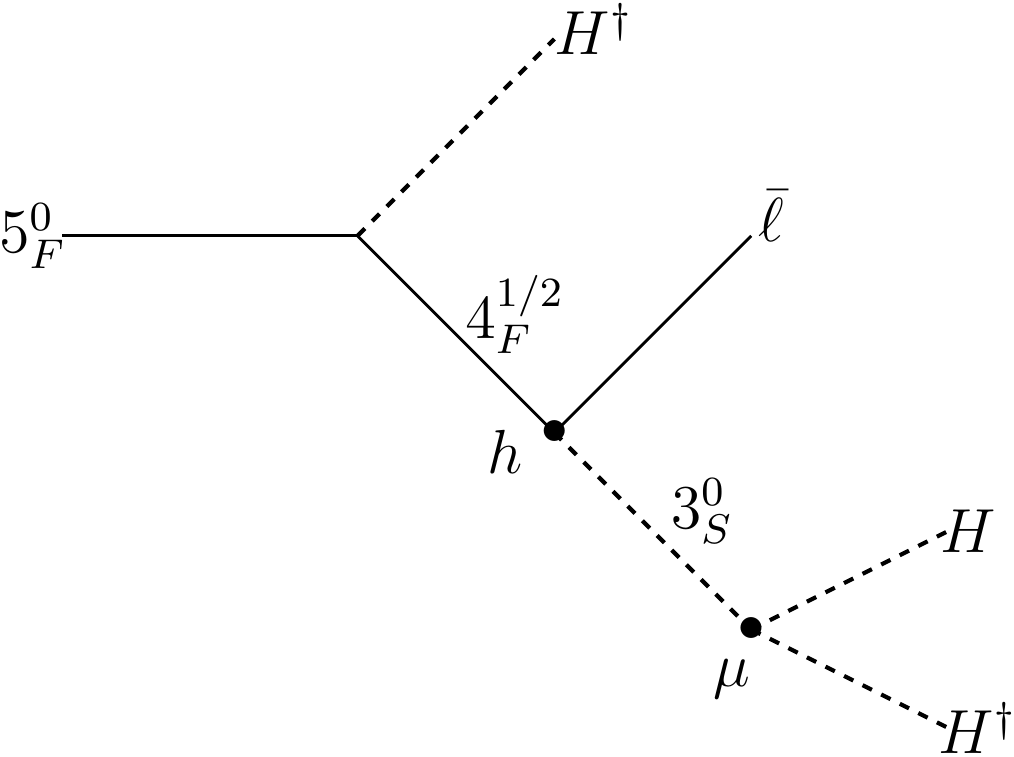}\\[3mm]
    \begin{minipage}{2cm}
      $(a)$
    \end{minipage}
    \hspace{5cm}
    \begin{minipage}{2cm}
      $(b)$
    \end{minipage}
    %\hspace{2cm}
    %
    
    \vspace{0.3cm}
    \includegraphics[scale=0.55]{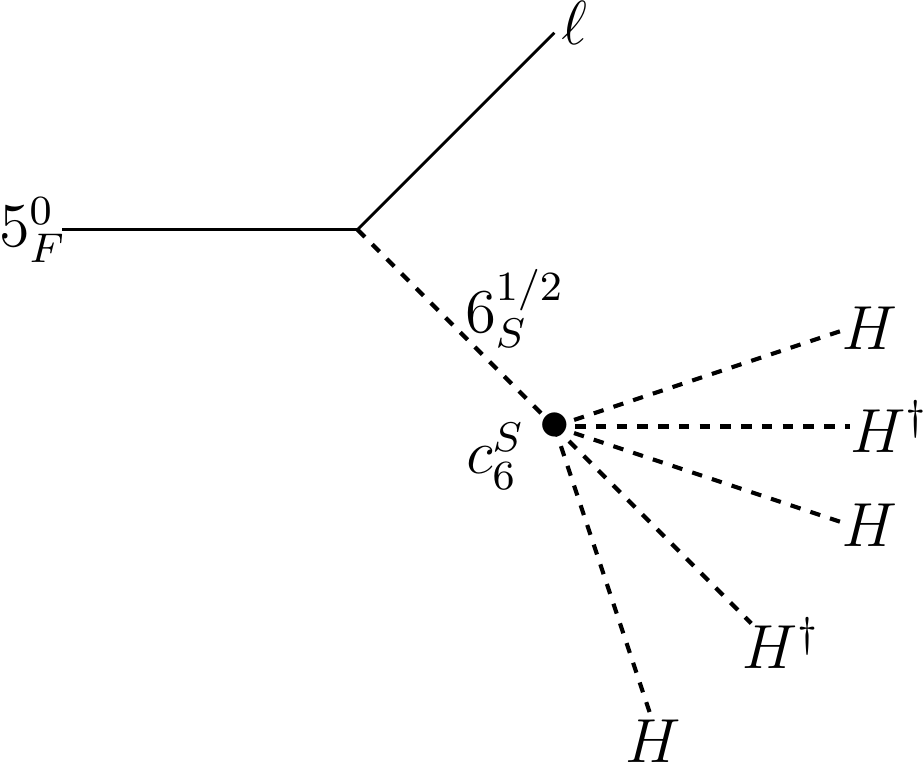}
    \hspace{2.5cm}
    \includegraphics[scale=0.55]{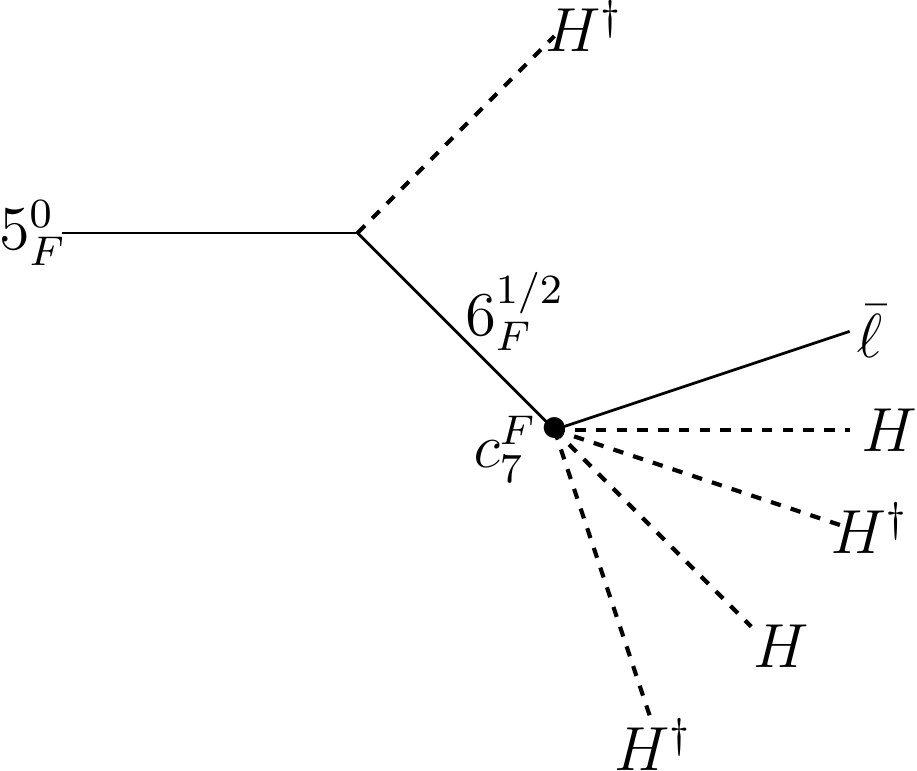}\\[3mm]
    \begin{minipage}{2cm}
      $(c)$
    \end{minipage}
    \hspace{5cm}
    \begin{minipage}{2cm}
      $(d)$
    \end{minipage}
    \caption{Upper panel: Tree level operators responsible for
      $\chi_\text{DM}\subset \boldsymbol{5}^0_F$ decay with a single
      (left) and multiple (right) messenger fields. Lower panel: Tree
      level operators for $\chi_\text{DM}\subset \boldsymbol{5}^0_F$
      decay for messenger fields not allowing renormalizable
      couplings.  Dots indicate $\mathbb{Z}_2$-breaking couplings.}
    \label{fig:tree-level}
  \end{figure}
  In single messenger scenarios containing
  $\boldsymbol{6}_{S,F}^{1/2}$, tree level processes involve dim=8
  operators (see fig. \ref{fig:tree-level}, diagrams $(c)$ and
  $(d)$). The decay processes will involve two distinct scales, namely
  $m$ and the scale at which the effective coupling $c_6^S$ or $c_7^F$
  is generated. This scale can be related with either the scale at
  which $\alpha_2=g_2^2/4\pi^2$ becomes non-perturbative or the Planck
  scale. Depending on the UV completion leading to $c_6^S$ or $c_7^F$ the
  former can be rather low, and thus in some instances
  $\chi_\text{DM}$ decay can be fast. If taken to be
  $M_\text{Planck}$, these processes cannot threaten DM stability. All
  in all, if one relies only on tree level processes one ends up with
  the conclusion that scenarios involving representations beyond the
  quartet (scalar or fermionic) are consistent with DM stability. This
  conclusion however is misleading since one-loop induced decay
  processes can always be written and can be fast enough to render
  $\chi_\text{DM}$ cosmologically unstable.
\item \textbf{One-loop level decay modes for $\boldsymbol{5}^0_F$}\\
  In this case as well one can distinguish single and multiple
  messenger scenarios, for the former the relevant representations are
  $\boldsymbol{R}= \boldsymbol{4}^{1/2}_S, \boldsymbol{6}^{1/2}_S$.
  These representations enable a $\mathbb{Z}_2$-breaking quartic
  coupling which renders $\chi_\text{DM}$ unstable.  The corresponding
  operators are shown in \ref{fig:one-loop-decay} (diagram to the
  left). These operators lead to different decay modes of which those
  with no Higgs condensation attachment are dominant, e.g.
  $\nu\,W^\pm\,W^\mp\,Z_L$. One can conclude that in one-loop neutrino
  mass models containing a $\boldsymbol{4}^{1/2}_S$, DM decays are
  fast and proceed via tree and one-loop level processes.
  %Thus in general is not cosmologically stable.

  Loop-induced decay processes are of more relevance in models that
  contain a $\boldsymbol{6}^{1/2}_S$. At the tree level these
  scenarios involve processes that lead to slow DM decay, and so seem
  consistent with cosmological stability. In the effective limit, the
  loop-induced processes are associated with dim=6 effective operators
  whose cutoff scale amounts (at most) to $m=10^9\,$~GeV. Thus, an
  estimation of the DM lifetime can be done from
  eq.~(\ref{eq:total-decay-width}) by rescaling by the loop
  factor. For processes such as $\nu\,W^\pm\,W^\mp\,Z_L$ the result
  reads:
  \begin{equation}
    \label{eq:lifetime-radiative-pure-scalar}
    \tau_\text{DM}\lesssim 7.1\times 10^8\,
    \left(\frac{10^{-2}}{\lambda}\right)^2
    \left(\frac{10^4\,\text{GeV}}{m_\text{DM}}\right)^5\,
    \left(\frac{m}{10^9\,\text{GeV}}\right)^4
    \,\text{sec.}\, .
  \end{equation}
  Which shows that although the tree level decay does not yield fast
  decays, the loop-induced processes lead to lifetimes that do not
  amount to $10^{26}\,$~ seconds, under ``reasonable'' parameter
  choices. This conclusion is inline with what was first pointed out
  in ref. \cite{Kumericki:2012bf}, showing the infeasibility of the
  R$\nu$MDM \cite{Cai:2011qr}.
  \begin{figure}
    \centering
    \includegraphics[scale=0.65]{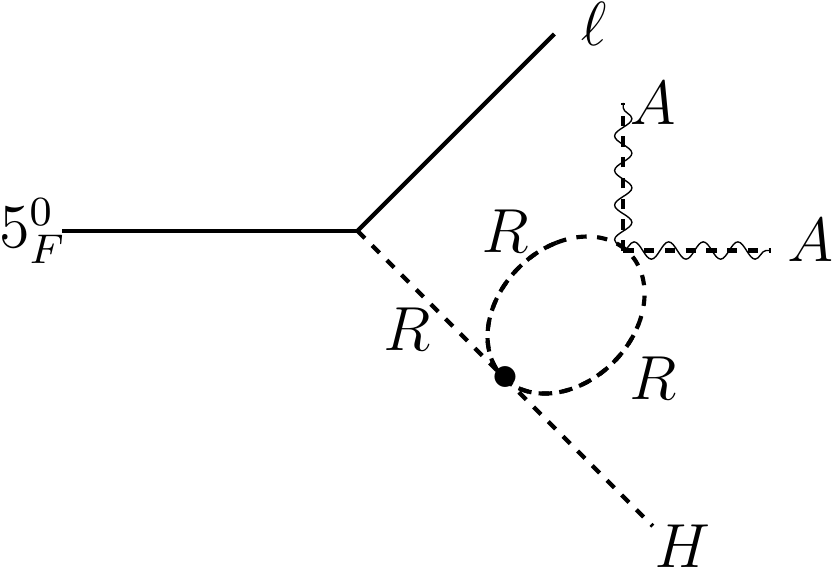}
    % diagram15.pdf
    \hspace{2.5cm}
    \includegraphics[scale=0.65]{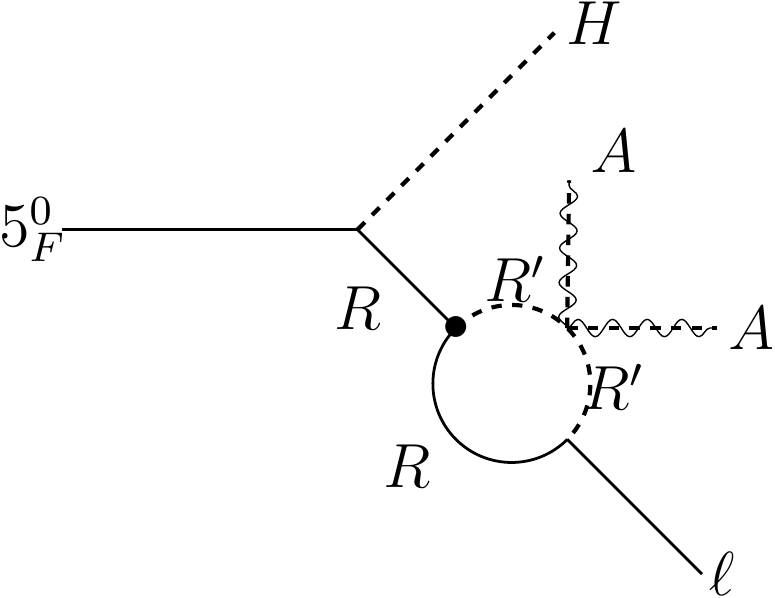}\\[2mm]
    %diagram8.pdf
    \begin{minipage}{1cm}
      \textbf{sl-i}
    \end{minipage}
    \hspace{6cm}
    \begin{minipage}{1cm}
      \hspace{5cm}{\textbf{fl-i}}
    \end{minipage}
    \caption{One-loop diagrams responsible for
      $\chi_\text{DM}\subset \boldsymbol{5}_F^0$ instability.  Dots
      indicate $\mathbb{Z}_2$-breaking couplings. Label \textbf{sl-i}
      refers to scalar loops, while \textbf{fl-i} to fermion
      loops. For the diagram to the left
      $\boldsymbol{R}=\boldsymbol{4}^{1/2}_S$ or
      $\boldsymbol{R}=\boldsymbol{6}^{1/2}_S$ (\textbf{sl-1} and
      \textbf{sl-2}, respectively), while for the diagrams to the
      right $(\boldsymbol{R}, \boldsymbol{R}^\prime)$:
      $(\boldsymbol{4}^{1/2}_F,\boldsymbol{3}^0_S)$,
      $(\boldsymbol{4}^{1/2}_F,\boldsymbol{5}^0_S)$,
      $(\boldsymbol{6}^{1/2}_F,\boldsymbol{5}^0_S)$,
      $(\boldsymbol{6}^{1/2}_F,\boldsymbol{7}^0_S)$ (\textbf{fl-1},
      \textbf{fl-2}, \textbf{fl-3} and \textbf{fl-4},
      respectively). The wiggling lines refer to $A\,A=H\,H^\dagger$
      for $\boldsymbol{R}=\boldsymbol{4}_{S,F}$, while to $A\,A=V\,V$
      ($V=W,Z,\gamma$) for $\boldsymbol{R}=\boldsymbol{6}_{S,F}$, as
      required by weak isospin conservation.}
    \label{fig:one-loop-decay}
  \end{figure}
  In the case of multiple messenger fields one finds Yukawa couplings
  of the type
  $\boldsymbol{R}^{1/2}_F\,\,\boldsymbol{R}^{-1/2}_F\,(\boldsymbol{R}^\prime)_S^0$,
  which break the $\mathbb{Z}_2$ accidental symmetry. Their presence
  allow the construction of operators as those shown in
  fig.~\ref{fig:one-loop-decay} (diagram to the right), that induce
  again processes such as $\chi_\text{DM}\to \nu\,W^\pm\,W^\mp\,Z_L$.
  The lifetime for these processes amounts to the value of the single
  messenger case in (\ref{eq:lifetime-radiative-pure-scalar}).
\end{itemize}
We now turn to the $\boldsymbol{7}_S^0$ DM scenario. As we have
already stressed, in the context of minimal DM this representation is
not consistent with stability even for
$\Lambda=M_\text{Planck}$. Thus, even if a one-loop neutrino mass
model does not involve degrees of freedom that can generate at the
renormalizable level the operator in
(\ref{eq:eff-7})---arguably---quantum gravity effects will generate it
and therefore $\varphi_\text{DM}\subset \boldsymbol{7}_S^0$ will decay
at a fast rate \cite{DiLuzio:2015oha}. It is however worth identifying
those extra representations that enable writing UV completions of this
effective operator, in particular because these results will enable in
turn the identification of one-loop neutrino mass models where
$\varphi_\text{DM}$ decays will be due to neutrino physics itself,
rather than due to a different type of physics.

\begin{figure}
  \centering
  \includegraphics[scale=0.6]{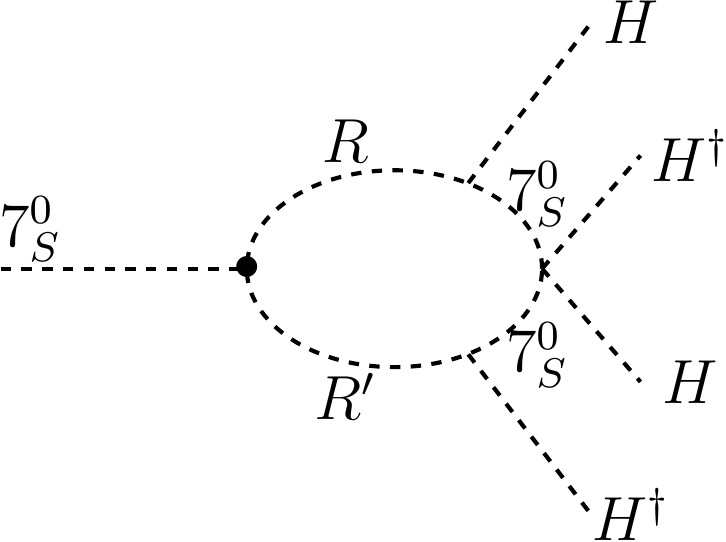}
  \hspace{2.5cm}
  \includegraphics[scale=0.6]{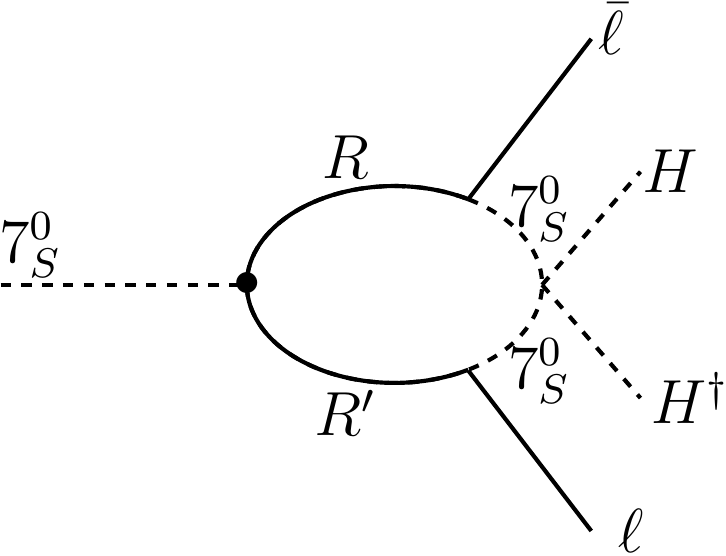}\\[2mm]
  \begin{minipage}{1cm}
    \textbf{sl-i}
  \end{minipage}
  \hspace{6cm}
  \begin{minipage}{1cm}
    \hspace{5cm}{\textbf{fl-i}}
  \end{minipage}
  \caption{Loop-induced operators responsible for
    $\varphi_\text{DM}\subset \boldsymbol{7}_S^0$ decays. Both
    diagrams involve three possible operators according to:
    $\boldsymbol{R}=\boldsymbol{R}^\prime=\boldsymbol{6}^{1/2},
    \boldsymbol{8}^{1/2}$
    and $\boldsymbol{R}=\boldsymbol{6}^{1/2}$ and
    $\boldsymbol{R}^\prime=\boldsymbol{8}^{1/2}$. The label
    \textbf{sl-i} refer to scalar loops, while \textbf{fl-i} to
    fermion loops. In both cases, \textbf{i}=1 for
    $\boldsymbol{R}= \boldsymbol{6}^{1/2}$, \textbf{i}=2 for
    $\boldsymbol{R}=\boldsymbol{6}^{1/2}$ and
    $\boldsymbol{R}^\prime=\boldsymbol{8}^{1/2}$ and \textbf{i}=3 for
    $\boldsymbol{R}= \boldsymbol{8}^{1/2}$. The dot indicates the
    $\mathbb{Z}_2$-breaking interaction.}
  \label{fig:seven-decay-loop-induced}
\end{figure}

As in the fermionic DM scenario, in this case one can identify various
$\mathbb{Z}_2$-breaking terms whose structure depends on the extra
representations present. We have found trilinear scalar couplings of
the type
$\boldsymbol{7}_S^0\,\boldsymbol{R}_S^{1/2}\,(\boldsymbol{R}_S^\prime)^{-1/2}$
and Yukawa couplings
$\boldsymbol{7}_S^0\,\boldsymbol{R}_F^{1/2}\,(\boldsymbol{R}_F^\prime)^{-1/2}$,
with
$\boldsymbol{R}=\boldsymbol{R}^\prime=\boldsymbol{6}, \boldsymbol{8}$
and $\boldsymbol{R}=\boldsymbol{6}$ and
$\boldsymbol{R}^\prime=\boldsymbol{8}$ (for both, fermions and
scalars). With these vertices one can then construct a certain number
of loop-induced processes that lead to fast $\varphi_\text{DM}$
decays. The different operators are shown in
fig. \ref{fig:seven-decay-loop-induced}. The different cases can be
sorted in single and multiple messenger scenarios, but due to our
universal mass simplification they lead to the same lifetime. Possible
decay modes for operators to the left in
fig.~\ref{fig:seven-decay-loop-induced} are $W^\pm_L\,W^\mp_L$, while
for those to the right $l^\pm\,l^\mp$. In the effective limit,
$m\gg m_\text{DM}$, the decay lifetime for processes induced by
operators as those shown in the diagram to the left in
fig.~\ref{fig:seven-decay-loop-induced} has been calculated as
\cite{DiLuzio:2015oha}:
\begin{equation}
  \label{eq:partial-width-7-effective}
  \Gamma_\text{DM}=\frac{857\,C_0^2}{441548\,\pi^5}\frac{g^2\,v^4}{m^2\,m_\text{DM}}\, ,
\end{equation}
with $C_0\simeq -0.0966$. The lifetime can then be estimated to be
\begin{equation}
  \label{eq:lifetime-seven-eff}
  \tau_\text{DM}\lesssim 5.9\times 10^3\,\left(\frac{m_\text{DM}}{10^4\,\text{GeV}}\right)\,
  \left(\frac{m}{10^9\,\text{GeV}}\right)^2\,\text{sec.}\, .
\end{equation}
For processes with leptons modes (induced by the operators to the right
in fig. \ref{fig:seven-decay-loop-induced}) the lifetime is larger, as
can be readily understood by realizing that the effective operator is
dim=6 rather than 5, as in the pure scalar case. Though larger, the
values are always well below what cosmological stability demands.
% -----------------
% Section 2
% -----------------
\section{UV completions of one-loop realizations of the Weinberg
  operator}
\label{sec:UV-completions}
In order to be model-independent one has to consider all possible
one-loop realizations of the lepton-number-breaking dim=5 operator.
The possibilities have been systematically sorted in
\cite{Bonnet:2012kz}, and diagrammatically their number reduces to four
independent diagrams as shown in fig. \ref{fig:one-loop-weinberg}.  We
start by assuming that one of the loop degrees of freedom is either
$R=\boldsymbol{5}_F^0$ or $R=\boldsymbol{7}_S^0$. As soon as this
internal representation is fixed the remaining representations are
fixed by $SU(2)\times U(1)_Y$ invariance. It is worth emphasizing that
in some instances the resulting representations allow for tree level
realizations of higher-order lepton-number-breaking operators (see
e.g. ref. \cite{Kumericki:2012bh}). Their contributions to neutrino
masses are however negligible provided the BSM fields have masses
above $\sim$ 3 TeV.
\subsection{The hypercharge-zero quintet case}
\label{sec:quintet}
\begin{figure}
  \centering
  \includegraphics[scale=0.7]{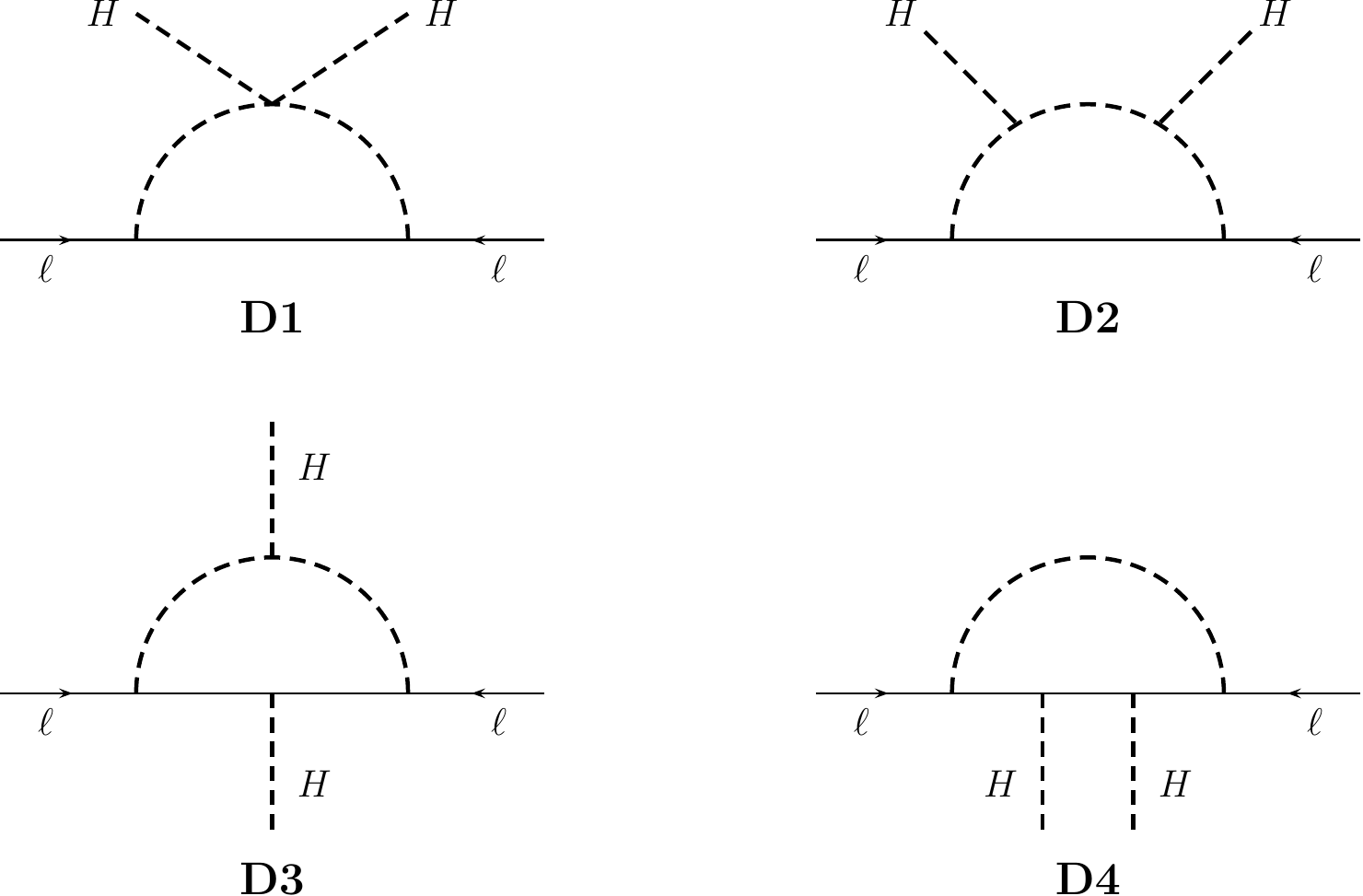}
  \caption{One-loop diagrams for UV realizations of the dim=5 Weinberg operator.}
  \label{fig:one-loop-weinberg}
\end{figure}

Results for the possible UV completions that contain a
hypercharge-zero fermionic quintet are shown in
tab. \ref{tab:quintet-results}. The different models can be readily
derived with the aid of the $SU(2)$ product decomposition
\begin{equation}
  \label{eq:5plet-product-decompositions}
  \boldsymbol{2}\otimes \boldsymbol{R} = \boldsymbol{R} \pm 1\, ,
  % \qquad
  % \boldsymbol{2}\otimes \boldsymbol{4} = \boldsymbol{3} \oplus \boldsymbol{5}\, ,
  % \qquad
  % \boldsymbol{2}\otimes \boldsymbol{6} = \boldsymbol{5} \oplus \boldsymbol{7}\, ,
\end{equation}
and thus rather than listing them one by one we discuss their generic
features.  A simple inspection to \textbf{D1} in
fig. \ref{fig:one-loop-weinberg}, shows that fixing
$\boldsymbol{5}_F^0$ two representations for the internal scalars are
possible, namely $\boldsymbol{4}_S^{1/2}$ and
$\boldsymbol{6}_S^{1/2}$. A diagram combining both is of course
possible too. In terms of the necessary number of representations, UV
completions based on \textbf{D1} define minimal scenarios. UV
completions based on \textbf{D2} require at least an additional scalar
representation, and in contrast to those based on \textbf{D1} are such
that the representation content always allow for two different
contributions to the effective neutrino mass matrix:
\begin{equation}
  \label{eq:nmm-D2}
  m_\nu= m_\nu^\text{\textbf{D2}} + m_\nu^\text{\textbf{D1}}\, ,
\end{equation}
where $m_\nu^\text{\textbf{D1}}$ refers to a contribution to the
neutrino mass matrix involving $\boldsymbol{4}_S^{1/2}$,
$\boldsymbol{6}_S^{1/2}$ or both. Scenarios based on \textbf{D3} require
one extra fermion and two scalar representations. The quantum numbers
demanded by gauge invariance are such that the neutrino mass matrix in
these cases have always the form
\begin{equation}
  \label{eq:nmm-D2}
  m_\nu= m_\nu^\text{\textbf{D3}} + m_\nu^\text{\textbf{D2}} 
  + m_\nu^\text{\textbf{D1}}\, ,
\end{equation}
regardless of the representations defining \textbf{D3}. UV completions
based on \textbf{D4} are of two types. Those for which the mass matrix
is entirely determined by \textbf{D4} and those that involve a
contribution arising from \textbf{D1}. The former case is obtained
whenever in addition to $\boldsymbol{5}_F^0$, there is a
$\boldsymbol{4}_F^{1/2}$ or a $\boldsymbol{6}_F^{1/2}$, or both
simultaneously. The latter, instead, requires in addition to
$\boldsymbol{4}_F^{1/2}$ or $\boldsymbol{6}_F^{1/2}$ an extra
fermionic representation that can be $\boldsymbol{3}_F^1$,
$\boldsymbol{5}_F^1$ or $\boldsymbol{7}_F^1$.
% ---------------------------------
% LNV minimal quintet DM table
% ---------------------------------
\begin{table}
  \centering
  \renewcommand{\arraystretch}{1.5}
  \setlength{\tabcolsep}{7pt}
  \begin{tabular}{|c|c|c|c|c|c|c|c|c|c|c|c|c|}
    \hline
    \multicolumn{13}{|c|}{\textbf{Fermion quintet radiative UV completed models}}
    \\\hline
    % row 1
   \multirow{2}{1cm}{\textbf{UV}} 
    & \multirow{2}{*}{$\boldsymbol{3}^1_F$} 
    & \multirow{2}{*}{$\boldsymbol{4}^{1/2}_F$} 
    & \multirow{2}{*}{$\boldsymbol{5}^1_F$}
    & \multirow{2}{*}{$\boldsymbol{6}^{1/2}_F$}
    & \multirow{2}{*}{$\boldsymbol{7}^1_F$} 
    & \multirow{2}{*}{$\boldsymbol{3}^0_S$}
    & \multirow{2}{*}{$\boldsymbol{4}^{1/2}_S$}
    & \multirow{2}{*}{$\boldsymbol{5}^0_S$}
    & \multirow{2}{*}{$\boldsymbol{6}^{1/2}_S$} 
    & \multirow{2}{*}{$\boldsymbol{7}^0_S$}
    %& \multirow{2}{*}{$\Lambda_\text{Non-pert}$}
    & \multicolumn{2}{c|}{\textbf{Decay}}\\\cline{12-13}
    &&&&&&&&&&&\textbf{Mode 1}&\textbf{Mode 2}\\\hline
    % row 2
    \textbf{D1} &  --  & --   &  --  & --
       & --   & --  & \cellcolor{Khaki1}{\ding{51}}   & --  & \cellcolor{DarkOliveGreen1}{\ding{51}}
       & -- & % $10^{17}$~GeV &
                                \cellcolor{Khaki1}{\bf sl-1} & \cellcolor{DarkOliveGreen1}{\bf sl-2}
    \\\hline
    % row 3
    \textbf{D2} & -- & -- & -- & -- 
    & -- & \ding{51} & \cellcolor{Khaki1}{\ding{51}} & \ding{51} & \cellcolor{DarkOliveGreen1}{\ding{51}}
    & \ding{51} & % $10^{14}$~GeV &
                                  \cellcolor{Khaki1}{\bf sl-1} & \cellcolor{DarkOliveGreen1}{\bf sl-2}
    \\\hline
    % row 4
    \textbf{D3} & -- & \cellcolor{Khaki1}{\ding{51}} & -- & \cellcolor{DarkOliveGreen1}{\ding{51}}
    & -- & \cellcolor{Khaki1}{\ding{51}} & \ding{51} & \cellcolor{LightBlue1}{\ding{51}} & \ding{51}
    & \cellcolor{DarkOliveGreen1}{\ding{51}} & % $5\times 10^8$~GeV &
                                       \cellcolor{Khaki1}{\bf fl-i} & \cellcolor{DarkOliveGreen1}{\bf fl-j}
    \\\hline
    \textbf{D4} & \ding{51} & \cellcolor{Khaki1}{\ding{51}} & \ding{51} & \cellcolor{Khaki1}{\ding{51}}
    & \ding{51} & \cellcolor{Khaki1}{\ding{51}} & \cellcolor{DarkOliveGreen1}{\ding{51}} 
    & \cellcolor{Khaki1}{\ding{51}} & \cellcolor{DarkOliveGreen1}{\ding{51}}
    & \cellcolor{Khaki1}{\ding{51}} & % $3\times 10^9$~GeV &
                                       \cellcolor{Khaki1}{\bf fl-i} & \cellcolor{DarkOliveGreen1}{\bf sl-j}
    \\\hline
  \end{tabular}
  \caption{One-loop UV completions for models with a fermion quintet defined by 
    diagrams \textbf{D1}-\textbf{D4} in fig. \ref{fig:one-loop-weinberg}. For
    each diagram and for one of the internal fermions fixed to be 
    $\boldsymbol{5}_F^0$, the remaining possible representations are 
    listed  according to the notation $\boldsymbol{R}_a^Y$
    (with $\boldsymbol{R}$ referring to the $SU(2)$ representation,
    $a=F,S$ and $Y$ to hypercharge). Representations not allowed by gauge 
    invariance are indicated by a dash, while allowed representations are indicated 
    by checkmarks. For the $\mathbb{Z}_2$-breaking decay modes we consider only 
    loop-induced processes. Representations that induce a decay mode 
    are indicated by colors. For example, in the first row 
    $\boldsymbol{4}_S^{1/2}$ induces an \textbf{sl-1} decay processes, 
    while $\boldsymbol{6}_S^{1/2}$ an \textbf{sl-2}. For the decay modes see 
    fig. \ref{fig:one-loop-decay}. In row 3 (\textbf{D3}), \textbf{fl-i} refers to
    \textbf{fl-1} and \textbf{fl-2}, while \textbf{fl-j} to \textbf{fl-3} and
    \textbf{fl-4}. The blueish cell refers to a representation ``shared'' by 
    modes:  \textbf{fl-2} and \textbf{fl-3}. In row 4 (\textbf{D4}), \textbf{fl-i}
    holds for i=1,\dots, 4 and \textbf{sl-j} for j=1,2.}
  \label{tab:quintet-results}
\end{table}

As regards the $\mathbb{Z}_2$-breaking decay modes, as can be seen in
tab.~\ref{tab:quintet-results} all possible models always involve
representations that enable writing decay modes as those discussed in
sec. \ref{sec:dm-decay-processes} (see
fig. \ref{fig:tree-level}-\ref{fig:seven-decay-loop-induced}). In
table \ref{tab:quintet-results} we have collected models according to
the category they belong to, the category defined by diagrams
\textbf{D1}-\textbf{D4}, and according to the decay mode they lead to
(we have specified only loop-induced processes). Representations
yielding a $\mathbb{Z}_2$-violating process can be identified from the
following color scheme: Yellowish (greenish) cells for representations
mean that those representations induce a $\mathbb{Z}_2$-breaking mode
1 (2).
\subsection{The hypercharge-zero septet case}
\label{sec:non-viable}
We now turn to the discussion of the scalar septet. As we have already
stressed, for this representation there is a loop-induced effective
dim=5 operator which even in the absence of an UV completion (in
``pure'' minimal DM) leads to fast DM decay. Despite that, here for
completeness we identify the different one-loop neutrino mass models
for which UV completions of that effective decay operator can be
written.

The main results in this case are summarized in
tab. \ref{tab:septet-results}.  The specification of the models in
these cases go along the same lines that in the quintet DM scenarios,
with additional $SU(2)$ product decompositions (e.g.
$\boldsymbol{2}\otimes \boldsymbol{7} = \boldsymbol{6} \oplus
\boldsymbol{8}$) given by eq.~(\ref{eq:5plet-product-decompositions}).
% given again But, in addition to the $SU(2)$ products
% in~(\ref{eq:5plet-product-decompositions}) other decompositions are
% needed, namely
% \begin{equation}
%   \label{eq:7plet-product-decompositions}
%   \boldsymbol{2}\otimes \boldsymbol{7} = \boldsymbol{6} \oplus \boldsymbol{8}\, ,
%   \qquad
%   \boldsymbol{2}\otimes \boldsymbol{8} = \boldsymbol{7} \oplus \boldsymbol{9}\, ,
%   \qquad
%   \boldsymbol{2}\otimes \boldsymbol{9} = \boldsymbol{8} \oplus \boldsymbol{10}\, .
% \end{equation}
With the aid of these product rules the different UV completions,
which we do not list, can be easily derived. Their main features are
the following. With one of the loop degrees of freedom fixed according
to $\boldsymbol{7}^0_S$, UV completions based on \textbf{D1} are of
two types, those involving $\boldsymbol{6}^1_F$ and those with
$\boldsymbol{8}^1_F$. Models relying on \textbf{D2} are of two types:
models where the mass matrix is entirely determined by
\textbf{D2}-type contributions and scenarios where the neutrino mass
matrix has the form:
\begin{equation}
  \label{eq:neutrino-mass-matrix-D2-case-septet}
  m_\nu= m_\nu^{\textbf{D2}} + m_\nu^{\textbf{D1}}\, .
\end{equation}
In \textbf{D3}-based scenarios the representations required imply in
all cases a neutrino mass matrix of the form
\begin{equation}
  \label{eq:neutrino-mass-matrix-D2-case-septet}
  m_\nu= m_\nu^{\textbf{D3}} + m_\nu^{\textbf{D2}} + m_\nu^{\textbf{D1}}\, ,
\end{equation}
Finally, for UV completions involving \textbf{D4} diagrams the
neutrino mass matrix involves only \textbf{D4} contributions.
% ---------------------------------
% LNV minimal septet DM table
% ---------------------------------
\begin{table}
  \centering
  \renewcommand{\arraystretch}{1.5}
  \setlength{\tabcolsep}{7pt}
  \begin{tabular}{|c|c|c|c|c|c|c|c|c|c|c|c|c|}
    \hline
    \multicolumn{13}{|c|}{\textbf{Scalar septet radiative UV completed models}}
    \\\hline
    % row 1
   \multirow{2}{1cm}{\textbf{UV}}
    & \multirow{2}{*}{$\boldsymbol{5}^0_F$} 
    & \multirow{2}{*}{$\boldsymbol{6}^{1/2}_F$} 
    & \multirow{2}{*}{$\boldsymbol{7}^0_F$}
    & \multirow{2}{*}{$\boldsymbol{8}^{1/2}_F$}
    & \multirow{2}{*}{$\boldsymbol{9}^0_F$} 
    & \multirow{2}{*}{$\boldsymbol{5}^1_S$}
    & \multirow{2}{*}{$\boldsymbol{6}^{1/2}_S$}
    & \multirow{2}{*}{$\boldsymbol{7}^1_S$}
    & \multirow{2}{*}{$\boldsymbol{8}^{1/2}_S$} 
    & \multirow{2}{*}{$\boldsymbol{9}^1_S$}
    %& \multirow{2}{*}{$\Lambda_\text{Non-pert}$}
    & \multicolumn{2}{c|}{\textbf{Decay}}\\\cline{12-13}
    &&&&&&&&&&&\textbf{Mode 1}&\textbf{Mode 2}\\\hline
    % row 2
    \textbf{D1} &  --  & \cellcolor{Khaki1}{\ding{51}}   &  --  & \cellcolor{DarkOliveGreen1}{\ding{51}}
       & --   & \ding{51}  & --   & \ding{51}  & --
       & \ding{51}
     %& $10^{13}$~GeV
    & \cellcolor{Khaki1}{\bf fl-1} & \cellcolor{DarkOliveGreen1}{\bf fl-3}
    \\\hline
    % row 3
    \textbf{D2} & \ding{51} & \ding{51} & \ding{51} & \ding{51} 
    & \ding{51} & \ding{51} & \cellcolor{Khaki1}{\ding{51}} & \ding{51} & \cellcolor{DarkOliveGreen1}{\ding{51}}
    & \ding{51} 
    % & $2\times 10^{10}$~GeV
    & \cellcolor{Khaki1}\textbf{sl-1} & \cellcolor{DarkOliveGreen1}{\textbf{sl-3}}
    \\\hline
    % row 4
    \textbf{D3} & \ding{51} & \cellcolor{Khaki1}{\ding{51}} & \ding{51} & \cellcolor{Khaki1}{\ding{51}}
    & \ding{51} & -- & \cellcolor{DarkOliveGreen1}{\ding{51}} & -- & \cellcolor{DarkOliveGreen1}{\ding{51}}
    & -- 
    %& $8\times 10^8$~GeV 
    & \cellcolor{Khaki1}{\bf fl-i} & \cellcolor{DarkOliveGreen1}{\bf sl-j}
    \\\hline
    % row 5
    \textbf{D4} & \ding{51} & \cellcolor{Khaki1}{\ding{51}} & \ding{51} & \cellcolor{DarkOliveGreen1}{\ding{51}}
    & \ding{51} & -- & -- & -- & --
    & -- 
    % & $4\times 10^7$~GeV 
    & \cellcolor{Khaki1}{\bf fl-1} & \cellcolor{DarkOliveGreen1}{\bf fl-3}
    \\\hline
  \end{tabular}
  \caption{One-loop UV completions for models with a scalar septet defined by 
    diagrams \textbf{D1}-\textbf{D4} in fig. \ref{fig:one-loop-weinberg}. For
    each diagram and for one of the internal scalars fixed to be 
    $\boldsymbol{7}_S^0$, the remaining possible representations are 
    listed  according to the notation $\boldsymbol{R}_a^Y$ (with $\boldsymbol{R}$ 
    referring to the $SU(2)$ representation, $a=F,S$ and $Y$ to hypercharge). 
    Representations not allowed by gauge invariance are indicated by a dash, 
    while allowed representations are indicated by checkmarks. For the 
    $\mathbb{Z}_2$-breaking decay modes we consider only loop-induced processes. 
    Representations that induce a decay mode are indicated by colors. For example, 
    in the first row $\boldsymbol{6}_F^{1/2}$ induces an \textbf{fl-1} decay 
    processes, while $\boldsymbol{8}_F^{1/2}$ an \textbf{fl-3}. For the different 
    decay modes see figure \ref{fig:seven-decay-loop-induced}. In row 3 (\textbf{D3}), 
    \textbf{fl-i} refers to \textbf{fl-1} and \textbf{fl-3}, while \textbf{sl-j}
    to \textbf{sl-1} and \textbf{sl-3}.}
  \label{tab:septet-results}
\end{table}

Regarding $\mathbb{Z}_2$-breaking modes, as shown in
tab.~\ref{tab:septet-results} all UV completions of one-loop
realizations of the Weinberg operator always involve a
$\mathbb{Z}_2$-violating decay mode. As we have pointed out in
sec. \ref{sec:dm-decay-processes}, compared with the fermion quintet
these processes lead to faster DM decay since they are related with
dim=5 decay operators.  The color scheme follows the same conventions
of tab. \ref{tab:quintet-results}: Yellowish (greenish) cells for
representations mean that those representations induce a
$\mathbb{Z}_2$-breaking mode 1 (2), with the different modes referring
to fig.~\ref{fig:seven-decay-loop-induced}.

\section{Pitfalls of DM from higher-order representations 
 and neutrino masses}
\label{sec:possible-caveats}
The different decay modes we have identified depend upon
neutrino-related parameters and $\mathbb{Z}_2$-breaking couplings. In
the estimations of lifetimes we have taken the former to their extreme
values, while keeping perturbative Yukawa couplings ($\left| Y_\nu\right|<1$). Thus,
if one takes the extra representations to be heavier the largest
neutrino mass will be below $m_\text{Atm}^\text{Exp}\simeq 0.05\,$~eV.
The only way that enables the increasing of the lifetime is through
couplings not related with neutrino physics that can be either
$\mathbb{Z}_2$-breaking or $\mathbb{Z}_2$-conserving. For
neutrino-related parameters taken to their extreme values, couplings
of order $10^{-11}$ or so (see
e.g. eq. (\ref{eq:lifetime-radiative-pure-scalar})) will lead to
lifetimes of order $10^{26}\,$~seconds.  Stability, however, is not
assured anymore by an accidental symmetry but from the smallness of a
parameter. Whether such small coupling is justified depends on the UV
completed model, e.g. it could be that in some UV realizations the
coupling is radiatively induced or generated from the spontaneous
symmetry breaking of a symmetry present in the UV completed theory (it
has a dynamical origin\footnote{For models of cosmological stable DM
  with dynamically generated small couplings see
  e.g. \cite{Sierra:2009zq}.}), case in which a small value would be
justified. One could as well argue that for $\mathbb{Z}_2$-breaking
couplings their smallness is technically ``natural'', in 't Hooft's
sense \cite{'tHooft:1979bh}, since tiny values increase the symmetry
of the Lagrangian. This statement, however, depends on their RGE
stability and therefore on their nature. For $\mathbb{Z}_2$-breaking
Yukawas, diagrams contributing to RGE running are proportional to the
Yukawa itself (or certain power) and so one expects their values to be
RGE stable. For scalar quartic couplings, instead, this is not
necessarily the case and strongly depends on the dimension of the
representations involved \cite{Hamada:2015bra}. Thus, in the absence
of an explicit UV completed model DM slow decays are probably better
justified if they arise from small Yukawas.

Note that even smaller couplings are required if one lowers the mass
scale of the messenger fields. The point is that with
$m\sim 10^9\,$~GeV any possible low-energy indirect probe
(e.g. lepton-flavor-violating or collider-related observables) is well
below current and future experimental sensitivities. Thus, testability
of these scenarios requires $m\ll 10^9\,$~GeV. For scales comparable
to those of DM the lifetimes are reduced by several order of
magnitudes and so smaller couplings, order $10^{-21}$ or so, are
required to achieve consistent lifetimes. Thus, in summary, in
neutrino mass models that contain extra representations as those we
have identified in the previous section (in addition to
$\boldsymbol{5}_F^0$ or $\boldsymbol{7}_S^0$, or both) one expects DM
to decay fast, with the decays determined by $\mathbb{Z}_2$-violating
operators (see
figs. \ref{fig:tree-level}-\ref{fig:seven-decay-loop-induced}). The
explicit breaking of the accidental $\mathbb{Z}_2$ symmetry renders DM
cosmologically unstable. Tuning of couplings is possible, but in that
case cosmological stability is unrelated with the mechanism underlying
minimal DM models.

\begin{figure}
  \centering
  \includegraphics[scale=1.]{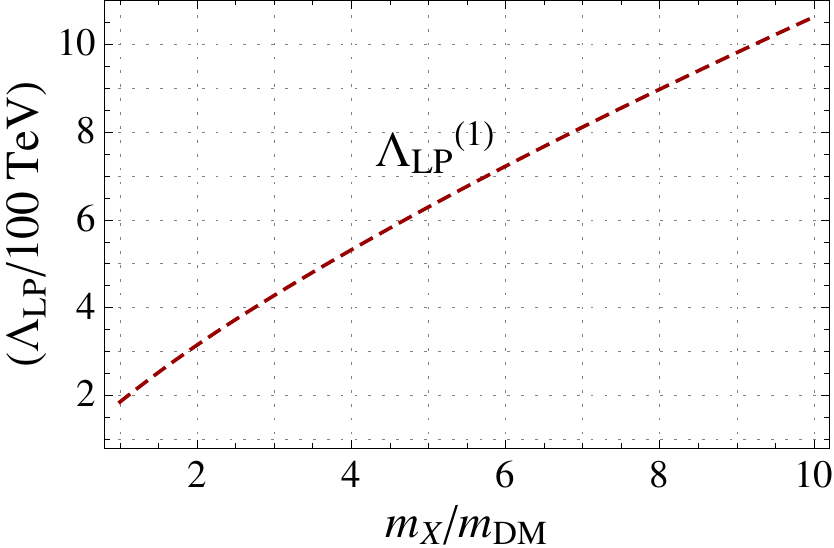}
  \hfill
  \includegraphics[scale=1.]{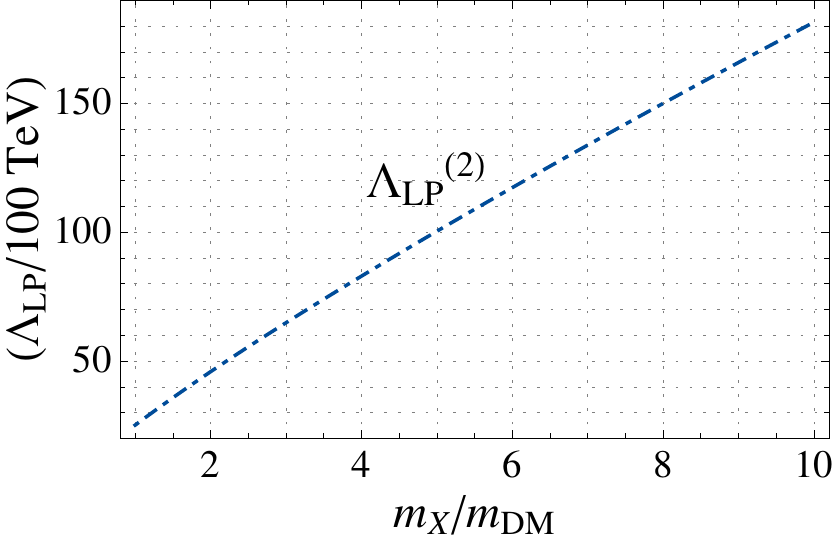}
  \caption{Energy scale at which $\alpha_2$ reaches a Landau pole as a
    function of $m_X/m_\text{DM}$ in the three-loop neutrino models of
    \cite{Ahriche:2015wha,Culjak:2015qja} ($m_X$ refers to the masses
    of the heavier fermion and scalars). The plot to the left is the
    result for the model in \cite{Ahriche:2015wha} with the DM mass
    fixed to 20 TeV and the singlet scalar ($\boldsymbol{1}_S^{1}$)
    mass to 500 GeV. The plot to the right is for
    \cite{Culjak:2015qja} with the DM mass fixed to 10 TeV and the
    second doublet mass to 1 TeV.  These results have been derived
    from two-loop RGEs and neglecting Yukawa contributions
    \cite{Machacek:1983tz} (see text for more details).}
  \label{fig:three-loop}
\end{figure}
A more plausible way of relating Majorana neutrino masses with
accidentally stabilized DM is through higher-order loops. In this
case, $SU(2)\times U(1)_Y$ invariance is less restrictive due to the
topologies of the possible diagrams. In the two-loop case, the
topologies allow for overall shifts in the hypercharge of the fields
flowing in the loops, while keeping $Y=0$ for the DM representation
(for the different topologies see ref. \cite{Sierra:2014rxa}). This
``hypercharge freedom'' might enable the construction of consistent
two-loop models. For three-loop models, as far as we are aware, two
examples have been pointed out \cite{Ahriche:2015wha,Culjak:2015qja}.
Ref. \cite{Ahriche:2015wha} considers a model with three
$\boldsymbol{7}_F^0$, one $\boldsymbol{7}_S^1$ and one
$\boldsymbol{1}_S^1$. The neutral component of the lightest
$\boldsymbol{7}_F^0$ is cosmologically stable due to an accidental
$\mathbb{Z}_2$ symmetry.  Ref. \cite{Culjak:2015qja}, instead,
considers a model with three copies of the $\boldsymbol{5}_F^0$, one
copy of the $\boldsymbol{7}_S^0$ and one copy of the
$\boldsymbol{5}_S^1$, in addition to a two-Higgs-doublet model. In
this case, the neutral component of the lightest $\boldsymbol{5}_F^0$
is cosmologically stable. Although consistent with DM stability,
neutrino physics data and low-energy lepton-flavor-violating
constraints, we have found that in both cases $\alpha_2=g_2^2/4\pi$
reaches a Landau pole at relatively low energy scales:
$\Lambda_\text{Landau-pole}^{(1)}\simeq 10\,m_X$,
$\Lambda_\text{Landau-pole}^{(2)}\simeq 10^2\,m_X$ (see
fig. \ref{fig:three-loop}), where $m_X$ refers to the masses of the
heavy representations which we have taken to be universal, for
simplicity.  This result has been derived by integrating two-loop RGEs
and neglecting Yukawa couplings contributions. Note that since in
three-loop models Yukawas are order one, their inclusion will change
the location of the Landau pole, with the precise value increasing or
decreasing depending on numerical details.

This result then shows that in these scenarios perturbativity is lost
at scales a couple orders of magnitude higher from the characteristic
energy scale of the models (taken to be $m_X$), with such behavior
being more pronounced for the model in \cite{Ahriche:2015wha}. In the
absence of a strongly coupled $SU(2)\times U(1)_Y$ sector at that
scale, ideally one should demand perturbativity to hold at least up to
the GUT scale, where one could expect that a new gauge sector assures
$\alpha_2<1$. Thus, arguably, this may be a drawback of these
approaches. The presence of multiple higher-order representations can
lead to non-perturbative scenarios at relatively low energy
scales. There is of course something one should bear in mind. One
could in principle understand these scenarios as effective limits of a
larger UV theory, in which it could be that by taken into account all
the degrees of freedom and intercations perturbativity could be
restored. Furthermore, in that theory it could be as well that the
$\mathbb{Z}_2$-violating couplings that we have identified as
responsible for DM fast decay are either absent or dynamically
suppressed, case in which DM will become cosmologically stable.
% -----------------
% Section 4
% -----------------
\section{Conclusions}
\label{sec:concl}
In this paper, we have presented a model-independent study of one-loop
neutrino mass models in which one of the loop messengers is a
hypercharge-zero fermion quintet or a hypercharge-zero septet. In the
absence of additional higher-order EW representations the neutral
component of the quintet is absolutely stable at the renormalizable
level, and cosmologically stable when the leading-order effective
operator responsible for its decay (dim=6) is included (assuming a
cutoff above $10^{15}\,$~GeV, see
fig. \ref{fig:lifetimes-minimal-DM-only}). We have shown that such
feature no longer holds in the presence of the extra degrees of
freedom that define the one-loop neutrino mass model, regardless of
the model. For completeness we have extended our analysis to include
the case of a hypercharge-zero scalar septet, although this
representation cannot be reconciled with cosmological stability even
in minimal DM scenarios.

We have systematically classified all such models and the
corresponding DM decay operators in each case. Our main findings are
summarized in tabs.~\ref{tab:quintet-results} and
\ref{tab:septet-results}, combined with the different DM decay
operators given in figs.~\ref{fig:one-loop-decay} and
\ref{fig:seven-decay-loop-induced}.  From these results our main
conclusion is that one-loop neutrino mass models that use minimal DM
representations are not consistent with DM stability, unless tiny
couplings---$\mathcal{O}\sim 10^{-21}-10^{-11}$ (with the value
depending on the highest scale of the corresponding neutrino mass
model)---or ad hoc symmetries are assumed. The former being
hard to reconcile in the absence of a dynamical mechanism assuring
such extreme small values, while the latter missing the whole point
behind higher-order EW representations\footnote{From the cosmological
  point of view, these representations are motivated for they provide
  stabilization without the need of additional symmetries.}.

We have commented on neutrino mass models variations based on
higher-order loops. We argued that in the two-loop case, models
consistent with DM stability might be possible writing due to the
``hypercharge freedom'' that the two-loop topologies offer. We have
stressed that these models and three-loop scenarios might be hard to
reconcile with perturbativity criteria. We have illustrated that for
two specific three-loop models for which we have found that $\alpha_2$
reaches a Landau pole at relatively low energy scales
(fig.~\ref{fig:three-loop}),
$\Lambda_\text{Landau-pole}\sim (10-10^2)m_X$, with $m_X$ the
characteristic scale of the model.
% ------------------------
% Section 3
% ------------------------
\section{Acknowledgments}
We would like to thank Sofiane Boucenna, Thomas Hambye, Alejandro
Ibarra, Igor Ivanov and Luca di Luzio for useful comments.  We
specially thank Michael Schmidt and Yi Cai for useful email exchanges
and for sharing with us results of their work \cite{Cai:2016jrl} prior
to its submission to arXiv. This work was supported by the ``Fonds de
la Recherche Scientifique-FNRS'' under grant number 4.4501.15. The
work of C.S. is supported by the ``Universit\'e de Li\`ege'' and the
EU in the context of the MSCA-COFUND-BeIPD project.
% ------------------------
% The bibliography
% ------------------------
%\bibliographystyle{references}

\vspace{0.7cm}
\noindent
\textbf{\large Note added in proof}\vspace{0.3cm}\\
While completing this work, two papers reaching conclusions similar to
ours appeared \cite{Cai:2016jrl,Ahriche:2016rgf}.
% Version v1 of \cite{Cai:2016jrl} studied 
% %
% Version v1 of
% \cite{Ahriche:2016rgf} (``Radiative Neutrino Mass via Both Minimal
% Dark Matter Candidates'') presented various one-loop models which were
% presumably consistent with minimal DM slow decays since they did not
% involve $\mathbb{Z}_2$-violating scalar couplings. Subsequently, the
% paper was updated (``A Critical Analysis of One-Loop Neutrino Mass
% Models with Minimal Dark Matter'') with a version in which it was
% shown that all one-loop neutrino mass models with minimal DM are not
% consistent with slow DM decay, since all of them allow for
% $\mathbb{Z}_2$-breaking couplings (scalar or Yukawa), hence reaching
% the same conclusion we have obtained here.
%
\bibliography{references}
\end{document}